# Quantum bit commitment protocol without quantum memory

Rubens Viana Ramos and Fábio Alencar Mendonça

rubens@deti.ufc.br          alencar@deti.ufc.br

Department of Teleinformatic Engineering, Federal University of Ceara, Campus do Pici, 725, C.P. 6007, 60755-640, Fortaleza-Ceará, Brazil

**Abstract**

Quantum protocols for bit commitment have been proposed and it is largely accepted that unconditionally secure quantum bit commitment is not possible; however, it can be more secure than classical bit commitment. In despite of its usefulness, quantum bit commitment protocols have not been experimentally implemented. The main reason is the fact that all proposed quantum bit commitment protocols require quantum memory. In this work, we show a quantum bit commitment protocol that does not require quantum memory and can be implemented with present technology.

**Key words:** Coherent states, quantum bit commitment.

## 1. Introduction

Bit commitment is one of the most important cryptographic protocols that can be used to realize, among others, coin tossing, zero-knowledge proofs and electronic voting. Due to its importance, it was one of the first cryptographic protocols that researchers tried to develop a quantum version [1-4], believing that a quantum bit commitment (QBC) protocol could be unconditionally secure. However, it was proved that it is impossible to construct an unconditionally secure QBC protocol using qubits encoded in single-photon pulses [5,6]. Basically, a bit commitment protocol has two stages: commit and unveil. In the commit stage Alice sends to Bob information representing (but not revealing) the bit value she chose. In the unveil stage Alice says to Bob the bit value she chose and Bob uses the information previously sent by Alice in order to check if Alice is saying the truth or lying. A bit commitment protocol is secure if it is binding and concealing. The protocol is binding if Alice can not change the value of the bit committed after the commit stage. On the other hand, a protocol is said to be concealing if Bob can not discover the bit value before the unveil stage. The insecurity of single-photon based quantum bit commitment was well explained by Lo and Chau in [5] and Mayer in [6]. Nevertheless, the fact that QBC is not perfectly secure does not mean that QBC has security similar to classical versions. In fact, limiting the power of the participants, it is possible to construct QBC protocols resistant to some types of attacks [7-9]. Recently, it was proposed a quantum bit string commitment protocol based on mesoscopic coherent states, QBSC_MCS [10]. Such protocol is more feasible with today technology. However, like all the others already proposed, it still requires quantum memory. In this work, we show another quantum bit/bit string commitment protocol using mesoscopic coherent states that does not require quantum memory and, hence, it can be fully implemented with present technology.

## 2. Quantum bit/bit string commitment protocol without quantum memory

Although the proposed bit string protocol can be extended to a large number of bits, he we are going to show its functioning only in a situation where Alice has to choose an option among three. The proposed protocol can be understood observing Fig. 1. In this figure, $B_1$ and $B_2$ are beam splitters, $R$ is a polarization rotator and PBS is a polarization beam splitter that resolves in the horizontal-vertical basis.

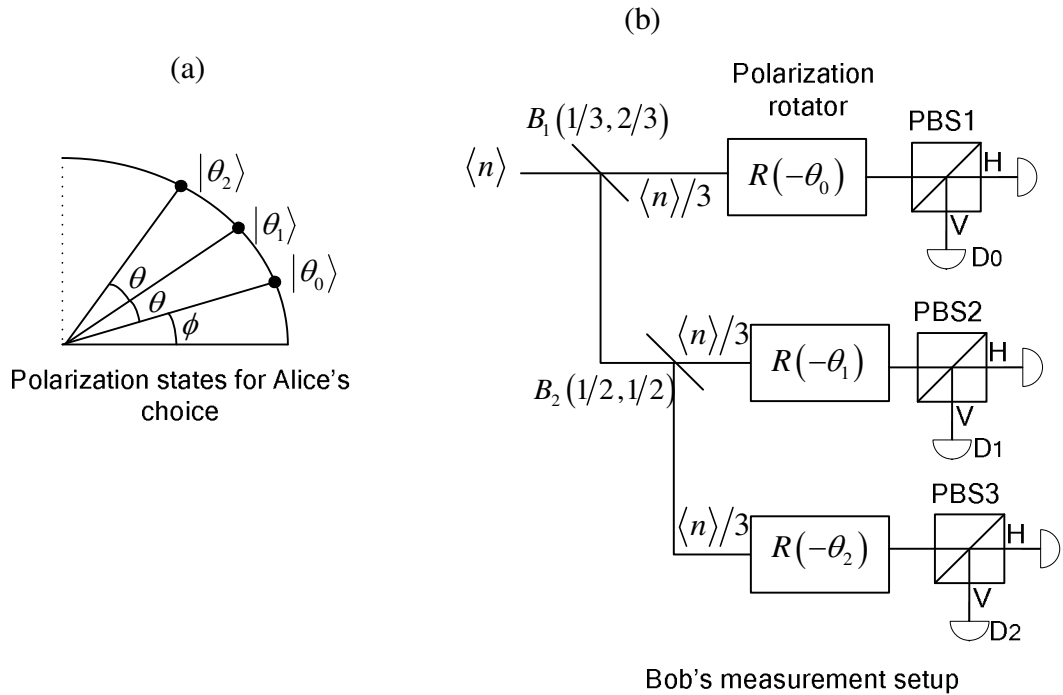

Fig. 1 - Quantum bit/bit string commitment protocol using coherent state. $B_1$ and $B_2$ are beam splitters, PBS is a polarization beam splitter and $R$ is a polarization rotator.

As can be seen in Fig. 1.a, each Alice's choice is coded in a polarization state represented by a two-mode coherent state:

$$|\theta_0\rangle = |\alpha\cos(\phi), \alpha\sin(\phi)\rangle \quad (1)$$

$$|\theta_1\rangle = |\alpha\cos(\phi+\theta), \alpha\sin(\phi+\theta)\rangle \quad (2)$$

$$|\theta_2\rangle = |\alpha\cos(\phi+2\theta), \alpha\sin(\phi+2\theta)\rangle \quad (3)$$

Here, we restrict ourselves to the cases where $\alpha$ is a real number ($\langle n \rangle = \alpha^2$), that is, only linear polarizations are considered. In Fig. 1.b, by its turn, one can see the apparatus used by Bob to measure the quantum state sent by Alice. According to the results of his measurements, in some cases Bob will be sure about the state sent by Alice (detections in two vertical outputs: $D_1$ and $D_2$, $D_1$ and $D_3$, or $D_2$ and $D_3$), in other cases he will be sure about the state not sent by Alice (detection in only one vertical output: $D_1$, $D_2$ or $D_3$) and, in the last, case he will not be sure about anything (none detection in the vertical outputs). All these situations are described in Table I, where the value '1' means detection.

| $D_1$ | $D_2$ | $D_3$ | States |
|---|---|---|---|
| 0 | 0 | 0 | $\{|\theta_0\rangle, |\theta_1\rangle, |\theta_2\rangle\}$ |
| 0 | 0 | 1 | $\{|\theta_0\rangle, |\theta_1\rangle\}$ |
| 0 | 1 | 0 | $\{|\theta_0\rangle, |\theta_2\rangle\}$ |
| 0 | 1 | 1 | $\{|\theta_0\rangle\}$ |
| 1 | 0 | 0 | $\{|\theta_1\rangle, |\theta_2\rangle\}$ |
| 1 | 0 | 1 | $\{|\theta_1\rangle\}$ |
| 1 | 1 | 0 | $\{|\theta_2\rangle\}$ |

Table I. Detections in Bob and his conclusions. Column 1 – vertical detections in Bob's setup. Column 2 – States possibly sent by Alice.

The proposed quantum bit string commitment protocol is as follows:

Commit stage:

1. Alice chooses one of the polarization states shown in Fig.1a and sends it to Bob.

2. Bob uses the apparatus shown in Fig. 1b to measure the quantum state sent by Alice and he stores the classical results obtained in his ordinary classical memory (memory of a computer).

Unveil stage:

1. Alice informs to Bob the quantum state sent.
2. Bob checks if Alice's information is in accordance with the results of his measurements. For example, Bob had detection in $D_2$. If Alice says to him that she sent $|\theta_2\rangle$ Bob will know she is lying.

The probability of Bob obtaining, unambiguously, the correct value of the state sent by Alice, before the unveil stage, is equal to the probability of Bob to identify, unambiguously, if the horizontal component of the unknown coherent state sent by Alice is one of the states $\{|\alpha\cos(\phi)\rangle, |\alpha\cos(\phi+\theta)\rangle, |\alpha\cos(\phi+2\theta)\rangle\}$. Alternatively, Bob could check if the vertical component of the unknown coherent state sent by Alice is one of the states $\{|\alpha\sin(\phi)\rangle, |\alpha\sin(\phi+\theta)\rangle, |\alpha\sin(\phi+2\theta)\rangle\}$. Using the setup proposed in [11], the upper-bound (using ideal photodetectors) for the probability of success, $P_{B1}$ for horizontal component and $P_{B2}$ for vertical component, is:

$$P_{B1} = \frac{1}{3}\left[\sum_{j=1}^{3}\prod_{k \neq j}\left(1-e^{-\frac{1}{\sqrt{2}}[\alpha_k - \alpha_j]^2}\right)\right] = \frac{1}{3}\left[1-e^{-\frac{\langle n \rangle}{\sqrt{2}}[\cos(\phi+\theta)-\cos(\phi)]^2}\right]\left[1-e^{-\frac{\langle n \rangle}{\sqrt{2}}[\cos(\phi+2\theta)-\cos(\phi)]^2}\right] +$$

$$\frac{1}{3}\left[1-e^{-\frac{\langle n \rangle}{\sqrt{2}}[\cos(\phi)-\cos(\phi+\theta)]^2}\right]\left[1-e^{-\frac{\langle n \rangle}{\sqrt{2}}[\cos(\phi+2\theta)-\cos(\phi+\theta)]^2}\right] +$$

$$\frac{1}{3}\left[1-e^{-\frac{\langle n \rangle}{\sqrt{2}}[\cos(\phi)-\cos(\phi+2\theta)]^2}\right]\left[1-e^{-\frac{\langle n \rangle}{\sqrt{2}}[\cos(\phi+\theta)-\cos(\phi+2\theta)]^2}\right] \quad (4)$$

$$P_{B2} = \frac{1}{3}\left[\sum_{j=1}^{3}\prod_{k\neq j}\left(1-e^{-\frac{1}{\sqrt{2}}[\alpha_k-\alpha_j]^2}\right)\right] = \frac{1}{3}\left[1-e^{-\frac{\langle n\rangle}{\sqrt{2}}[\sin(\phi+\theta)-\sin(\phi)]^2}\right]\left[1-e^{-\frac{\langle n\rangle}{\sqrt{2}}[\sin(\phi+2\theta)-\sin(\phi)]^2}\right]+$$

$$\frac{1}{3}\left[1-e^{-\frac{\langle n\rangle}{\sqrt{2}}[\sin(\phi)-\sin(\phi+\theta)]^2}\right]\left[1-e^{-\frac{\langle n\rangle}{\sqrt{2}}[\sin(\phi+2\theta)-\sin(\phi+\theta)]^2}\right]+$$

$$\frac{1}{3}\left[1-e^{-\frac{\langle n\rangle}{\sqrt{2}}[\sin(\phi)-\sin(\phi+2\theta)]^2}\right]\left[1-e^{-\frac{\langle n\rangle}{\sqrt{2}}[\sin(\phi+\theta)-\sin(\phi+2\theta)]^2}\right] \quad (5)$$

On the other hand, the probability of Alice cheating Bob, $P_A$, is:

$$P_A = p_{000} + \frac{1}{2}(p_{001} + p_{010} + p_{100}) \quad (6)$$

$$P_{000} = \frac{1}{3}e^{-\frac{\langle n\rangle}{3}\sin^2(\theta)}\left[e^{-\frac{\langle n\rangle}{3}\sin^2(\theta)} + 2e^{-\frac{\langle n\rangle}{3}\sin^2(2\theta)}\right] \quad (7)$$

$$p_{001} = p_{100} = \frac{1}{3}e^{-\frac{\langle n\rangle}{3}\sin^2(\theta)}\left[2 - e^{-\frac{\langle n\rangle}{3}\sin^2(2\theta)} - e^{-\frac{\langle n\rangle}{3}\sin^2(\theta)}\right] \quad (8)$$

$$p_{010} = \frac{2}{3}e^{-\frac{\langle n\rangle}{3}\sin^2(2\theta)}\left[1 - e^{-\frac{\langle n\rangle}{3}\sin^2(\theta)}\right] \quad (9)$$

In (6), $p_{000}$ is the probability of none vertical detection, $p_{001}$ ($p_{010}$, $p_{100}$) is the probability of vertical detection only in $D_2$ ($D_1$, $D_0$). If Bob has none vertical detection, then the probability of Alice cheating him without being discovered is 1. If Bob has one vertical detection in $D_i$, then he can detect a lie if Alice says during the unveil stage that she sent $|\theta_i\rangle$. Hence, the probability of Alice cheating Bob without being discovered, in this case, is ½. At last, if Bob has detections in two vertical outputs, then the probability of Alice cheating him without being discovered is 0, since in this case Bob will be sure about the state sent by Alice.

For an ideal quantum bit string commitment protocol, one should have $P_A=P_B=0$. In practice, one can accept $P_A$ and $P_B$ larger than zero and to establish a hard penalty for that one that is caught cheating. Observing (4)-(9) one can observe that increasing the mean photon number, $P_A$ tends to zero and $P_B$ tends to one. On the other hand, decreasing the mean photon number, $P_A$ tends to one and $P_B$ tends to zero. Hence, it is not possible to make simultaneously both equal to zero. An example is shown in Fig. 2, where it was used $\phi=\pi/6.7$ and $\theta=\pi/10$.

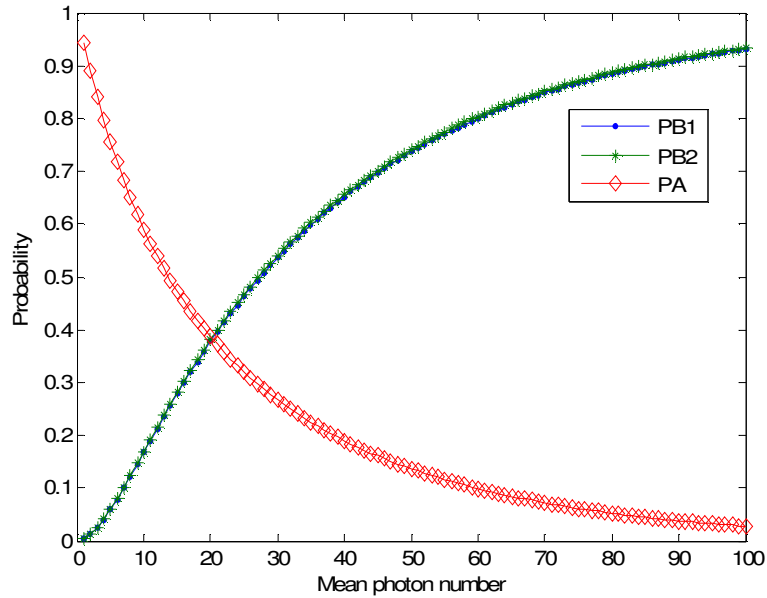

Fig. 2 - Success probabilities for Alice's and Bob's attacks versus mean photon number.

In Fig. 2, for $\langle n \rangle=20$, one finds $P_A \approx 0.386$, $P_{B1} \approx 0.3772$ and $P_{B2} \approx 0.3808$. We are going to use these values along the rest of the paper. The advantage of the parameter's values used is the fact that Bob's success probability is almost independent of his choice, thus he can measure the horizontal or vertical component. Therefore, Alice and Bob can use coherent states with mean photon number equal to 20 in order to run a quantum bit commitment protocol where they have less than 40% of probability to cheat each other. Even having the same probability of success than Bob for her attack, the protocol is not fully fair from Alice's point of view. If she is caught cheating, she will get a hard penalty. On the other hand, Bob will never be caught cheating because he just makes the measurements that are expected to be done according to the protocol. Thus, it is reasonable to decrease $P_B$ without increasing $P_A$. However, this can not be achieved by changing the physical parameters. In order to achieve this goal, we propose a slight modification in the quantum bit commitment protocol:

Commit stage:
1. Alice sends to Bob two quantum states. The pairs $\{|\theta_0\rangle|\theta_1\rangle,|\theta_1\rangle|\theta_0\rangle,|\theta_2\rangle|\theta_0\rangle\}$ represent bit 0 while the pairs $\{|\theta_0\rangle|\theta_2\rangle,|\theta_1\rangle|\theta_2\rangle,|\theta_2\rangle|\theta_1\rangle\}$ represent bit 1.
2. Bob uses the apparatus shown in Fig. 1b to measure the quantum states sent by Alice and he stores the classical results obtained.

Unveil stage:
1. Alice informs to Bob the quantum states sent.
3. Bob checks if Alice's information is in accordance with the results of his measurements.

If Alice wants to cheat Bob, she lies only about one of the states, hence, keeping the same parameters used before ($\phi=\pi/6.7$, $\theta=\pi/10$ and $\langle n\rangle=20$), her probability of success is $P_A \approx 0.386$. On the other hand, in order to get the bit value without Alice's permission, Bob has to guess the correct state twice and, hence, his probabilities of success are $P_{B1} \approx 0.1422$ and $P_{B2} \approx 0.145$.

## 3. Discussions

In a practical realization of the protocol, some cares must be taken into account. Firstly, Alice and Bob shall measure the total loss in their setups and in the link between them in order to correct the mean photon number that Alice shall use. The loss in Alice's and Bob's setups can be estimated from the knowledge of the loss present in common optical devices. The critical point is the loss in the fiber link between Alice and Bob. In fact, a dishonest Bob can try to cheat Alice making her to believe that the loss is larger than the real value, what makes her to increase the mean photon number. Bob can easily do this by placing a fiber spool in his setup before Alice's loss measurement (using optical time-domain reflectometer - OTDR) and taking it off before starting the QBC protocol. In this case, assuming the parameters values used before, the pulse arriving at Bob would have a mean photon number larger than 20. This increases the probability of Bob cheating Alice. In this case, if the mean photon number required by Bob is too large, Alice can deny it and does not play a bit commitment with Bob. This implies that Alice will run a bit commitment only with Bobs that are inside of a secure distance. It is worth to remind that Bob can not use a beam splitter to simulate a larger loss in the link, because Alice can measure the fiber length (measuring the photon time of flight) and calculate the total loss. The comparison between the measured valued and the calculated one using fiber attenuation of 0.27dB/km can reveal the use of a beam splitter.

On the other hand, a dishonest Alice can try to cheat Bob making him to believe that the loss is lower than the real value. This is a hard task to Alice since, before Bob's loss measurement, she would have to change the optical fiber (or at least a piece) by another one shorter (following a straight line) or with lower attenuation coefficient. The second point is the polarization rotation during optical fiber propagation. This can be controlled by using polarization correction schemes [12,13].

## 4. Conclusions

It was proposed a quantum bit commitment protocol, using coherent states, whose implementation does not rely on quantum memories and, hence, it can be implemented, with today technology. Furthermore, quantum version of other protocols based on bit commitment, like zero knowledge systems, can also be implemented.